# Ostwald Growth Rate in Controlled Covid-19 Epidemic Spreading as in Arrested Growth in Quantum Complex Matter

**Antonio Bianconi** [1,2,3,*], **Augusto Marcelli** [1,4], **Gaetano Campi** [1,2] **and Andrea Perali** [1,5]

1. Rome International Centre Materials Science Superstripes RICMASS via dei Sabelli 119A, 00185 Rome, Italy;
2. Institute of Crystallography, CNR, via Salaria Km 29. 300, Monterotondo Stazione, Roma I-00016, Italy gaetano.campi@ic.cnr.it
3. National Research Nuclear University MEPhI (Moscow Engineering Physics Institute), 115409 Moscow, Russia
4. INFN - Laboratori Nazionali di Frascati, 00044 Frascati (RM), Italy; Augusto.Marcelli@lnf.infn.it
5. School of Pharmacy, Physics Unit, Università di Camerino, 62032 Camerino (MC), Italy; andrea.perali@unicam.it

\* Correspondence: antonio.bianconi@ricmass.eu

**Abstract:** Here, we focus on the data analysis of the growth of epidemic spread of Covid-19 in countries where different policies of containment were activated. It is known that the growth of pandemic spread at its threshold is exponential, but it is not known how to quantify the success of different containment policies. We identify that a successful approach gives an arrested phase regime following the Ostwald growth, where, over the course of time, one phase transforms into another metastable phase with a similar free energy as observed in oxygen interstitial diffusion in quantum complex matter and in crystallization of proteins. We introduce the s factor which provides a quantitative measure of the efficiency and speed of the adopted containment policy, which is very helpful not only to monitor the Covid-19 pandemic spread but also for other countries to choose the best containment policy. The results show that a policy based on joint confinement, targeted tests, and tracking positive cases is the most rapid pandemic containment policy; in fact, we found values of 9, 5, and 31 for the success s factor for China, South Korea, and Italy, respectively, where the lowest s factor indicates the best containment policy.

**Keywords**: Covid-19, epidemic control policies, Biophysical approach, Ostwald arrested growth, big data analysis, success s factor, best epidemic containment policy

## 1. Introduction

While theories on epidemic spread in complex systems were developed in the last few years [1–6], the rapid Covid-19 epidemic spread in the strongly interconnected human world community in 2019 and 2020 provides an extraordinary challenge for the science of complex systems. Many researchers all around the world are currently engaged in the quantitative determination of the Covid-19 pandemic spread in almost all countries [7,8]. Open-source data provided by official health agencies are the essential tool for the scientific community to understand the type of diffusion process and determine the mathematical laws of the diffusion of the virus. Fortunately, few countries, e.g., China, South Korea, Italy, Singapore, and others provided verified data and chose the policy of transparency and containment of contagion through isolation of the population and other restrictive measures on the movement of people. Our data analysis of the pandemic growth curves was carried over about 28 days from the onset $t_0$ where the exponential growth rate characterized by the doubling time, called $T_d$, assumes the value $T_{d0}$ of about two days in the three different countries.

## 2. Results and Discussions

Considering the large database now available, it is possible to analyze the dynamics of the epidemic spread in different countries. The evolution of the epidemic spread shows the universality of the complex growth curves, pointing to the occurrence of multiple phases with transitions in a time-evolving process out of equilibrium. The first phase is the "pre-threshold phase" (PTP), showing random scattered events. This phase is characterized by a limited number of infected people which could cover very different laps of time in different sites depending on the local space topology, human density, public transport, social habits, and frequency of face-face interactions at short distance.

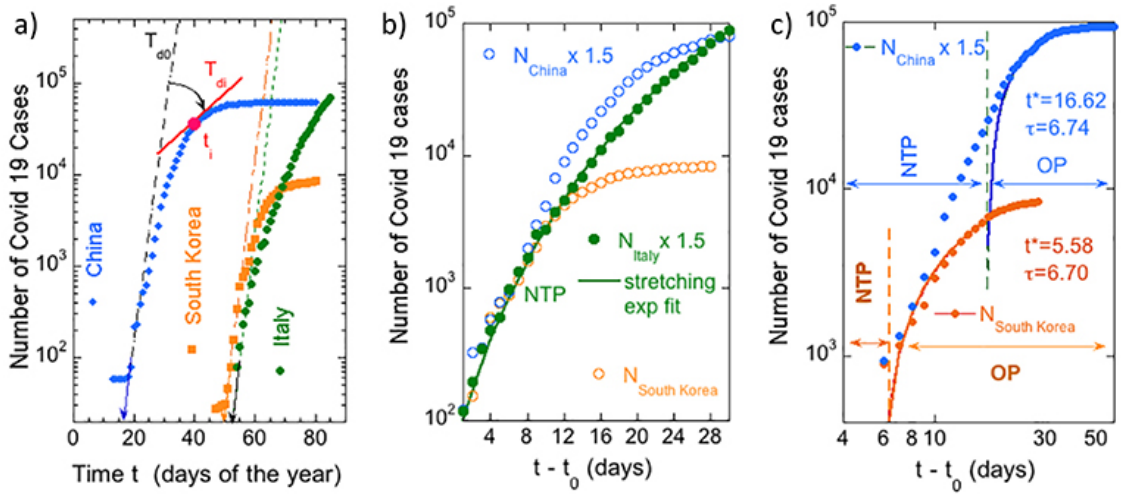

**Figure 1.** Number of contagion cases present in one day of 2020 in China (blue), South Korea (orange), and Italy (green). The curve of each country is characterized by the day $t_0$ where the transition of the system from the "below threshold" regime to the exponential regime characteristic of the pandemic spread which is indicated by a straight line in the semi-log plot in panel (**a**). At each time $t_i$, we extracted the time-dependent doubling time $T_{di}$ by fitting the data over a five-day period centered at the selected time $t_i$ as indicated by the red line in panel (a). The efficiency of the containment policy is indicated by the curvature of experimental curves. The curvature of the growth curves changes with time. Panel (**b**) shows the three growth curves in a semi-log scale in a common time scale using $t_0$ shown in panel (a) for each country as the starting time. The counts of the Italy and China cases are multiplied by 1.5 to overlap all curves in the near-threshold phase (NTP). The growth curve for pandemic spread in Italy is fitted with a stretched exponential. Panel (**c**) shows the growth curves in China and South Korea which show the arrested Oswald growth in the log–log scale after a time t*, which is typical of the arrested or frustrated growth in a complex potential landscape.

In Figure 1, we report the number of positive Covid-19 cases after the critical time $t_0$ for the onset of epidemic spread. The data show the growth regime in the second, third and fourth phases. The second phase starts on a day $t_0$ where the growth for $t > t_0$ shows the typical exponential law of the pandemic spread. This behavior is similar in the three countries considered and it is characterized by a doubling time $T_{d0}$ of about two days ($1.5 < T_d < 2$). If the process remains free, without introducing containment measures, the number of people infected would, therefore, double in less than three days. Under these conditions, about 10 million people would be infected in a three-week period as indicated by the straight lines in Figure 1a. Figure 1b shows the three different curves on the same time scale using $t_0$ of each curve as a zero time for the explosion of the epidemic spread.

The pandemic control policies put in place by Italy, China, and South Korea managed to modify the exponential growth. In the third regime, the "near-threshold phase" (NTP), common in all curves, is where the starting exponential regime shows a deceleration of the exponential growth. In the NTP regime, the growth curves show an exponential behavior with an increasing doubling time. The growth process of the Covid-19 in the near-threshold phase (NTP) of the containment measures was found in the Italian growth curve over a large time range shown in Figure 1b. We show that it does not follow the standard logistic curve or the Kolmogorov law, but a stretching exponential.

The fourth phase, called Ostwald phase (OP), can be identified in the log–log plot of data in Figure 1c where the growth curve of South Korea (China) is fitted from $t^* = 5.58$ d to 30 d (from $t^* = 16.62$ d to 60 d) after the threshold by using the formula of the Ostwald growth rate.

$$N(t - t^*) = C\{1 - e^{-(t-t^*)/\tau}\} \cdot (t - t^*)^\gamma. \tag{1}$$

The fit by the Ostwald formula (Equation (1)) of both curves of China and South Korea pandemic growth in Figure 1c in the arrested Ostwald phase (OP) show the same half time $\tau = 6.7$ d.

The South Korea and China growth curve were fitted by the Ostwald growth formula (Equation (1)) with a characteristic time scale $\tau$ of the Ostwald regime of 6.7 days, which indicates that the same time scale $2\tau = 13.4$ days is needed to reach the pandemic saturation level. Therefore, it is possible that the time scale $2\tau = 14$ days is a characteristic time scale to suppress the pandemic spreading by targeted test, mobile phone tracking, and geo-localization used in South Korea and China. The Ostwald arrested or frustrated growth phenomenon which we observed in the Ostwald phase is typically observed in diffusion processes in complex systems [9,10], as found by our research group in quantum materials [11–13]. Indeed, this trend is typically observed in non-homogeneous complex systems with "domain walls" or "topological solitons" that slow down the growth, with a function having a doubling time $T_d$ that changes continuously [14], which shows, that over the course of time, one phase transforms into another metastable phase with a free energy similar to that observed in the processes of crystallization of proteins [15] and metal phosphates [16].

To quantify the success of the containment policies, we show in Figure 2 the value of the doubling time T_d as a function of time, for China, blue curve (panel a), Korean, orange curve (panel b), and Italy, green curve (panel c), over a period of time covering both NTP and OP phases. An increase in doubling time denotes an improvement in the efficacy of the adopted strategy. It is clear that, at present, the South Korea policy is the most effective. Looking at the experimental behavior, the Ostwald growth is confirmed by the presence of kinks or steps in all curves, which indicate the occurrence of successive phase transitions from a metastable phase to another with a similar free energy, which are signatures of the dynamical inhomogeneity of the process.

The increasing doubling time indicates that the count $N$ of infections per day is less than what would have been observed in the case of non-governed exponential growth. The observed increasing value of $T_d$ as a function of time in the NTP and OP regimes are shown to follow an empirical exponential law,

$$T_d(t) = A\, e^{t/s}, \qquad (2)$$

measuring the slowdown of the rate of growth of the infection both in the near-threshold phase and in the Ostwald arrested phase. The success of the containment policy is measured by the shortest possible time to reach the day where there are no infected people per day. This state corresponds to an infinite doubling time $T_d$ achieved in China and in South Korea.

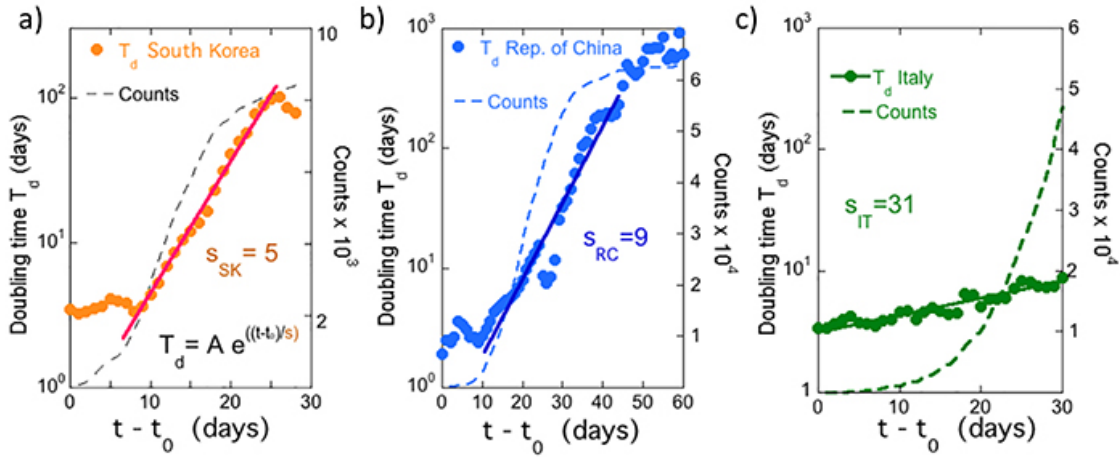

**Figure 2** Evolution of the doubling time in the three countries: (**a**) South Korea, (**b**) China, and (**c**) Italy. The figure shows the occurrence of steps in all curves, a phenomenon we associate with the occurrence of transitions to successive different phases. The "s factor" is the parameter obtained by exponential fit of the curves $T_d(t)$ of the doubling time vs. time over a period of days longer than 14 days $T_d(t) = A\, e^{t/s}$. The average period of time to get the s factor in this work was 40, 15, and 30 days for China, Korea, and Italy respectively. The results show that our success s factor probes the speed of the containment measures to reach the epidemic saturation point. A small success s factor measures the success of the confinement policy to keep the epidemic spread under control in a short time. The speed of the Italy policy is about six times slower that the Korea policy.

As can be seen, even if to a different extent, the three approaches were successful, whereby the different impact of the strategies can be immediately recognized. South Korea seems to have been able to reach the goal in a shorter time. This trend was quantified with the s factor, which must be as small as possible in order to reach the goal of zero people infected per day in the shortest possible time.

We call this the success s factor of the strategy to stop the pandemic spread, which was introduced in Reference [17]. The evolution of the doubling time reported in Figure 2 for the three different countries is fitted by an exponential functional dependence $T_d(t) = A\, e^{t/s}$ over a period of time longer than the five days used to calculate $T_d$. The average period of time to get the s factor was 40, 15, and 20 days for China, South Korea, and Italy, respectively. The success s factors for the three countries are s = 9 for China, s = 5 for South Korea, and s = 31 for Italy.

The emerging question to answer is why in South Korea the arrested phase started before the Chinese one. We can reasonably hypothesize that the protocol of positive cases tracing in South Korea started quickly. South Korea made over 250,000 tests corresponding to ~0.5% of its population, quickly identifying a large number of asymptomatic infections that were placed in quarantine [7], and tracking movements of positive cases is clearly correlated with the arrested growth rate. Moreover, only space confinement of the population is not enough; in fact, in the last 10 days since the first experimental report of 15 March [17], the success factor in Italy shows practically no changes, and in Italy the growth rate is still in the near-threshold phase (NTP) 30 days after the threshold, while the arrested Ostwald regime has not started. This may be due to a lack of mobile phone tracking being activated, although strict space confinement measures were taken.

## 3. Conclusions

In conclusion, our interdisciplinary biophysical research provides a new application of mathematical data analysis developed in the new field of quantum complex matter physics [11–13] to epidemic viral infection dynamics in humans.

We present a comparison of the growth of epidemic spread of Covid-19 in countries where a different policy of containment was activated. Our approach introduces the quality "s" success factor to provide a quantitative evaluation of the speed and efficiency of policies for arresting epidemic spreading. Our approach is based on the study of the physical evolution of pandemic spread, which shows the universality of the complex growth curve with multiple phases and transitions vs. time within the framework of out-of-equilibrium processes. In this framework, the results of our data analysis and, in particular, the introduction of the success s factor are not influenced by the large indeterminacy in the number of infected people and the lack of information on asymptomatic cases which indeed may hardly emerge.

For Covid-19, it is known that, at the threshold of the pandemic spread, the growth is exponential with a doubling time $T_{d0}$ of about two days. However, it was not known how to quantify the different policies addressed to stop epidemic growth curves. Our analysis clearly shows that South Korea and Chinese approaches were more effective than the Italian one, probably due to prompt action based on tracking positive cases. We show that, once data are shifted to the same time origin, as in Figure 1b and 1c, there is a large difference in the time lapse for the onset of the Ostwald arrested phase for different control strategies but when the Ostwald phase starts it can reach its goal in 14 days.

The Korean approach could also be applied in other countries, and the results show that governments and regional stakeholders must promote and make possible the use of this practice. This study strongly suggests to extend the mobile tracking approach in Italian regions and all over the world, in particular when the overall number of contagious persons is limited and the spreading process is in the pre-threshold phase (PTP) or in the near-threshold growth phase. At the present stage, numerous experts and institutions are in favor of targeted testing and mobile phone tracking of positive cases. People should be able to request free tests for themselves since the limitation of tests to only people with symptoms is unfortunately not an effective strategy, despite suggested by the World Health Organization (WHO) [18].

We think that the choice of WHO was made in the initial phase of the new epidemic phenomenon when data on the speed of Covid-19 epidemic spread and age-dependent lethality in different countries with different average life-long expectation were not known. The application of the success s factor can now provide a very fast method to help WHO to modify their prescriptions based on a quantitative measure of different velocities of different policies addressed to arrest the Covid-19 pandemic, which was not available before. We can freely provide the computer code we developed if requested to help non-profit organizations maintain control over this rapidly evolving epidemic spread. The s factor can be used to measure the efficiency of the needed confinement of infected people in order to cut interpersonal contact for the duration of quarantine of individuals during the peak of the epidemic phenomenon.

Finally, the results presented here show the capability of applying and tuning strategies in any country, ruling out the option of any "non-containment" strategy that exposes citizens to unacceptable risks. The main result of our work led to the introduction of the success s factor for a quantitative measure of policies to arrest the epidemic spread. Tracking positive cases with targeted tests represents the most rapid pandemic control policy; in fact, it can be easily arrested in a short time $2\tau = 14$ days as shown in both China and Korea, or it can be used to keep the epidemic spread under the threshold phase, thereby stopping the explosion of exponential growth as shown in both Singapore and Israel.

**Author Contributions:** The authors contributed equally to the conceptualization, methodology, and investigation while preparing the article.

**Funding:** This research was funded by Superstripes-onlus.

**Conflicts of Interest:** The authors declare no conflicts of interest.